\documentclass[a4paper,11pt]{article}
\pdfoutput=1 

\usepackage{jcappub} 

\usepackage[T1]{fontenc} 

\title{\boldmath Testing the cosmic distance duality relation using Type Ia supernovae and radio quasars through model-independent methods }

\author[a]{Fan Yang}
\author[a,1]{Xiangyun Fu \note{Corresponding author: xyfu@hnust.edu.cn}}
\author[b,2]{Bing Xu \note{Corresponding author: xub@ahstu.edu.cn}}
\author[c]{Kaituo Zhang}
\author[a]{Yang Huang}
\author[a]{Ying Yang}

\affiliation[a]{Department of Physics, Key Laboratory of Intelligent Sensors and Advanced Sensor Materials, Hunan University of Science and Technology, Xiangtan, Hunan 411201, China}
\affiliation[b]{School of Electrical and Electronic Engineering, Anhui Science and Technology University, Bengbu, Anhui 233030, China}
\affiliation[c]{Department of Physics, Anhui Normal University, Wuhu, Anhui 241000, China}

\abstract{In this work, we perform a cosmological-model-independent test on the cosmic distance duality relation (CDDR) by comparing the angular diameter distance (ADD) obtained from the compact radio quasars (QSOs) with the luminosity distance (LD) from the Pantheon Type Ia supernovae (SNIa) sample. The binning method and Artificial Neural Network (ANN) are employed to match ADD data with LD data at the same redshift, and three different parameterizations are adopted to quantify the possible deviations from the CDDR. We initially investigate the impacts of the specific prior values for the absolute magnitude $M_{\rm B}$ from SNIa and the linear size scaling factor $l$ from QSOs on the CDDR test, demonstrating that these prior values introduce significant biases in the CDDR test. To avoid the biases, we propose a method independent of $M_{\rm B}$ and $l$  to test CDDR, which treats the fiducial value of a new variable $\kappa\equiv10^{M_{\rm B} \over 5}\,l$ as a nuisance parameter and then marginalize its impact with a flat prior in the statistical analysis. The results show that the CDDR is consistent with the observational data, and QSOs can serve as a powerful tool for testing the CDDR independent of cosmological models.}

\begin{document}
\maketitle
\flushbottom

\section{Introduction}
The cosmic distance duality relation (CDDR) is a fundamental relation in modern cosmology~\cite{Etherington1993}, which relates the luminosity distance (LD)  $D_{\rm L}(z)$ and angular diameter distance (ADD) $D_{\rm A}(z)$ through the identity equation ${D_{\rm L}}={D_{\rm A}}{(1+z)}^{2}$, where $z$ is the cosmological redshift. This relation relies on three fundamental assumptions: the space-time is described by the metric theory, the light travels along the null geodesics between the source and the observer, and the photon number is conserved. As a fundamental relation, the CDDR has undoubtedly been applied in various research fields of astronomy, such as the large-scale distribution of galaxies and the uniformity of the Cosmic Microwave Background (CMB) temperature~\cite{Aghanim2020}, as well as the gas mass density distribution and temperature distribution of galaxy clusters~\cite{Cao2011,Cao2016}. In astronomical observations, any violation of the CDDR suggests the presence of new physics or unaccounted errors in the observational data. Therefore, it is necessary to conduct reliable testing on CDDR.

The CDDR test is usually conducted using a parametric method. Three different forms can be used for the parametrization, namely $\eta(z)=1+\eta_0z$, $\eta(z)=1+\eta_0z/(1+z)$, and $\eta(z)=1+\eta_0\ln(1+z)$, where the $\eta_{0}$ indicates the possible violation from the CDDR. Considering the advantages of the $\eta(z)$ expression, such as manageable one-dimensional phase space and good sensitivity to observational data~\cite{Holanda2011}, the parameterizations of $\eta(z)$ listed above were used to test the validity of the CDDR. Some studies have been devoted to testing the validity of CDDR by comparing LD data from observations of Type Ia supernovae (SNIa), HII galaxies, or gamma-ray bursts with the various ADD data from the X-ray plus Sunyaev-Zeldovich (SZ) effect and the gas mass fraction measurements in galaxy clusters~\cite{Uzan2004,DeBernardis2006,Holanda2010,Holanda2011,Holanda2012,Li2011,Nair2011,Meng2012,Yang2013,Santos2015,Hu2018,daSilva2020,Bora2021}. The results indicate that within various redshift ranges, the CDDR is consistent with astronomical observations~\cite{Avtidisgous2010,Stern2010,Holanda2012a,Holanda2013,Liao2011,Holanda20171,Fu2011,Fu2017,Liang2013,Liao2016,Xu2022,Zheng2020}. In addition, an issue may be that it is difficult to obtain LD and ADD measurements from astronomic observation at the same redshifts. To solve this problem, several methods have been proposed in some literature. Using the galaxy cluster samples~\cite{Bonamenteet2006,DeFilippis2005} and SNIa data, Holanda {\it et al.}~\cite{Holanda2010} and Li {\it et al.}~\cite{Li2011} selected the closest one through a selection criterion ($\Delta z=|z_{\rm ADD}-z_{\rm SNIa}|<0.005$) for CDDR test. To minimize statistical errors that could arise from utilizing only a single SNIa data point from all those that meet the selection criteria, Meng {\it et al.}~\cite{Meng2012}, did not use the closest measurement, but instead used a binning method to bin the available data that meets the selection criterion to derive LD.

SNIa and compact radio quasars (QSO) measurements play important roles in constraining cosmological parameters. It is worth noting that the LD derived from SNIa observations is dependent upon its peak absolute magnitude $M_{\rm B}$, which is assumed to be a constant value, unaffected by other variables. Recently, efforts have been made to derive the value of $M_{\rm B}$ from a cosmological observations~\cite{Camarena2020,Kumar2022,Dinda2023}. Various values of $M_{\rm B}$ have been determined by combining SNIa data, such as Pantheon, with other observational datasets, including CMB observations, cosmic chronometer data related to the Hubble parameter, and baryon acoustic oscillations (BAO). A discrepancy in the absolute magnitude of SNIa, calibrated by Cepheids, was observed between $z\leq0.01$ and $z>0.01$~\cite{Camarena2021,Camarena20201}. Recent studies~\cite{Kazantzidis2021,Kazantzidis2020} have also indicated a potential weak evolution of $M_{\rm B}$. In addition, due to the negligible dependence of the compact structure sizes of intermediate-luminosity quasars on the source luminosity and redshift, these quasars, with very-long-baseline interferometry (VLBI) observations, are potentially promising standard rulers~\cite{Cao2018}. The ADD obtained from these QSO observations depends on the linear size scaling factor $l$. The constraints on $l$ have been made using both cosmological-model-dependent  and cosmological-model-independent methods~\cite{Cao2017,Cao20171,Cao2019}, and the value of $l$ varies slightly depending on the observational data and the cosmological model.

Obviously, if the exact values of $M_{\rm B}$ and $l$ are not determined by astronomical observations, the specific prior values for $M_{\rm B}$ and $l$ may potentially introduce biases into the constraints on cosmological parameters. Furthermore,  the independence of the CDDR test, which relies on the prior values of $M_{\rm B}$ and $l$, may be questionable, as the values of $M_{\rm B}$ and $l$ are obtained from specific cosmological models. It is worth noting that the measurement error of a single SNIa or QSO measurement is not dependent on the parameter $M_{\rm B}$ or $l$, thus, theoretically, one can use marginalization methods to eliminate $M_{\rm B}$ or $l$ parameters during statistical analysis. In view of this, it is important to further study the impacts of specific prior values for $M_{\rm B}$ or $l$ on CDDR test, and to develop new testing methods that are not dependent on $M_{\rm B}$ and $l$, which can improve the reliability of CDDR testing. This is also the main motivation for us to carry out this research.

In this work, we perform the CDDR test by comparing the LD derived from Pantheon SNIa data with the ADD from QSO data. The binning method and Artificial Neural Network (ANN) are used to match the SNIa data with the QSO data at the same redshifts. We first investigate the impacts on the CDDR test by considering the specific value of $M_{\rm B}$ and $l$ to derive the LD and ADD. The results indicate that the priors of $M_{\rm B}$ and $l$ may induce significant biases in the CDDR test. To avoid these biases, we combine $M_{\rm B}$ and $l$ into a new variable $\kappa$, defined as $\kappa\equiv10^{M_{\rm B} \over 5}\,l$, and consider it as a nuisance  parameter with a flat prior in statistical analysis, thereby marginalizing its impact on CDDR test. Therefore, all of the quantities used in the CDDR test come directly from observations, meaning that the absolute magnitudes from SNIa and the linear size scaling factor from QSO measurements do not need to be calibrated. We demonstrate that CDDR is consistent with the observed data, and the parametric method of testing CDDR is independent of specific cosmological models.

\section{Data and Methodology}

\subsection{Data}
To verify the validity of CDDR, two kinds of cosmic distances are usually required: LD ($D_{\rm L}$) and ADD ($D_{\rm A}$). The LD in this work are obtained from the Pantheon SNIa observational data~\cite{Scolnic2018}, which includes 1048 data points from the Pan-STARRS1 Deep Survey in the redshift range of $0.01 < z < 2.26$. The distance modulus from the Pantheon compilation is calibrated from the SALT2 light-curve fitter by applying the Bayesian Estimation Applied to
Multiple Species with Bias Corrections method to determine the nuisance parameters and account for the distance bias corrections, such as $\mu = m_{\rm B}-M_{\rm B}$, where $m_{\rm B}$ is the observed peak apparent magnitude in the rest-frame B-band. Recently, some research has been focused on the possible evolution of $M_{\rm B}$ with redshift. The CMB constraint on the sound horizon forecasts that $M_{\rm B}\sim -19.4\,{\rm mag}$ using an inverse distance ladder~\cite{Dinda2023}, while the approximation from SH0ES gives that $M_{\rm B}\sim -19.2\,{\rm mag}$~\cite{Camarena2020}. Hence, we firstly investigate the impacts of different prior values of $M_{\rm B}$ on the CDDR test.  In this
work, we consider two specific priors of $M_{\rm B}$ derived from different observational data sets
within various redshift ranges: (a) $M_{\rm B}^{\rm D20}={-19.23\pm0.0404\,{\rm mag}}$, obtained from SNIa observation within the relatively low redshift range of $0.023 < z < 0.15$ by Camarena and Marra~\cite{Camarena2020} in $\Lambda$CDM, through a de-marginalization of the SH0ES determination~\cite{Reid2019} (hereafter referred to as $M_{\rm B}^{\rm D20}$); and (b) $M_{\rm B}^{\rm B23}={-19.396\pm0.016\,{\rm mag}}$, obtained by combining SNIa observations with BAO observations~\cite{Dinda2023} (hereafter referred to as $M_{\rm B}^{\rm B23}$). Taking into consideration the observational uncertainty of $M_{\rm B}$, the error bar on $\mu$ can be represented as $\sigma_{\mu}=\sqrt{\sigma_{M_{\rm B}}^2+\sigma_{m_{\rm B}}^2}$. The relation between the LD $D_{\rm L}$~\cite{Zhou2019} and the distance modulus $\mu$ can be expressed as
\begin{equation}
\mu(z)=5\log_{10}(D_{\rm L}(z))+25\,,
\end{equation}
and the uncertainty in $D_{\rm L}$ can be obtained from the equation
\begin{equation}
\sigma_{D_{\rm L}}={{\ln{10}}D_{\rm L}\sigma_{\mu}\over5}\,.
\end{equation}

The angular size-distance relationship of QSO is utilized for cosmological inference, originally proposed by Kellermann~\cite{Kellermann1993}, who attempted to obtain the deceleration parameter using VLBI observations of 79 compact radio sources at 5 GHz. Subsequently, Gurvits~\cite{Gurvits1994} extended this method and tried to study the dependence of the observed characteristic sizes of 337 Active Galactic Nuclei (AGNs) at 2.29 GHz on luminosity and redshift~\cite{Preston1985}. In the following analysis,  the angular size $\theta$ of the radio source is refined in Ref.~\cite{Gurvits1994} using the visibility modulus $\Gamma=S_{\rm c}/S_{\rm t}$, which can be expressed as $\theta={2\sqrt{-\ln{\Gamma}\ln2}\over\pi{B}}$, where $B$ is the interferometer baseline measured in multiple of wavelengths, and $S_{\rm c}$ and $S_{\rm t}$ are the correlated flux density and total flux density, respectively. The linear size $l_{\rm m}$ of compact structures in radio sources, the intrinsic luminosity $L$, and the redshift $z$ of the background source supply  the following relationship,
\begin{equation}
\label{lm}
l_{\rm m}=lL^{\beta}(1+z)^{n}\,,
\end{equation}
where $l$ represents the linear size scaling factor, describing the apparent distribution of radio brightness within the core, $\beta$ and $n$ are used to quantify the possible ``angular size-luminosity'' and ``angular size-redshift'' relations, respectively. Moreover, for a cosmological rod with intrinsic length, the relation of the angular size-redshift can be expressed as~\cite{Sandage1988}
\begin{equation}
\label{theta}
\theta(z)={l_{\rm m}\over{D_{\rm A}(z)}}\,,
\end{equation}
where $\theta(z)$ is the observed angular size measured by VLBI techniques. Combining Eq.~\ref{lm} and Eq.~\ref{theta}, the angular diameter distance $D_{\rm A}(z)$ can be written as
\begin{equation}
D_{\rm A}(z)={lL^{\beta}(1+z)^{n}\over{\theta(z)}}\,.
\end{equation}

Recently, Cao, {\it et al.} found that the linear size scaling factor is almost independent on redshift and intrinsic luminosity ($|n|\simeq10^{-3}$, $|\beta|\simeq10^{-4}$)~\cite{Cao2017,Cao20171}. The sample of 120 intermediate-luminosity radio quasars within redshift range of $0.4 < z < 2.8$ selected in~\cite{Cao2017} has been widely used in various cosmological studies~\cite{Cao2018,Liu2022,Ma2017,Qi2017}. The ADD obtained from the QSO samples has already been used to test the CDDR along with the LD obtained from HII galaxies and supernovae~\cite{Liu2021,He2022}, and to infer the value of the Hubble constant $H_0$ together with the unanchored luminosity from supernovae data~\cite{Liu2023}.

The value of the linear size scaling factor $l$ can be constrained to $l=11.19\pm1.64{\,\rm pc}$~\cite{Cao20171} (hereafter referred to as $l^{\rm C17}$) in the flat $\Lambda$CDM model with  WMAP9 observations. Then, in the manner of an independent study on cosmological model, Cao {\it et al.} obtained $l={10.86\pm1.58}{\,\rm pc}$~\cite{Cao20171}  by using 36 Hubble data points, some of which were inferred from 30 cosmic chronometers~\cite{Ding2015,Moresco2015,Moresco2016}, while the rest were derived from 6 BAO measurements~\cite{Zheng2016}.
Furthermore, Cao {\it et al.} obtained the more accurate value of $l$~\cite{Cao2019} (hereafter referred to as $l^{\rm C19}$) by using 41 Hubble data points, some of which were inferred from 31 passively evolving galaxies~\cite{Moresco2015,Jimenez2003,Simon2005,Stern2010,Chuang2012,Zhang2014}, while the rest were derived from 10 BAO measurements~\cite{Gaztanaga2009,Blake2012,Xu2013,Samushia2013,Delubac2013,Font-Ribera2014,Delubac2015}, which is $l^{\rm C19}={11.04\pm0.40}\,{\rm pc}$.
The values of $l$ provided by different observation data exhibit slight variations. Consequently, the prior values of $l$ may potentially induce bias in testing the CDDR. In this work, we consider the values of the linear size scaling factors $l^{\rm C17}$ and $l^{\rm C19}$, which are calibrated using methods that depend on and  do not depend on the cosmological model, respectively, to investigate their impacts  on the CDDR test.

\subsection{Binning method}
To test the validity of the CDDR, a straightforward method is to compare the ADD and LD from different observations at the same redshift. Due to the lack of observational ADD and LD data at the same redshift, we bin the SNIa data points satisfying the selection criterion $\Delta z=|z_{\rm ADD}-z_{\rm SNIa}|<0.005$, as proposed in the literature~\cite{Holanda2010,Li2011,Liao2016}. This method, known as the binning method, can be used to avoid statistical errors caused by using only one SNIa data point among those satisfying the selection criterion and has been employed in discussing the CDDR test in the cited Ref.~\cite{Meng2012,Wu2015}. Here, we take the inverse variance weighted average of all selected data. To avoid correlations among the individual CDDR tests, we select the SNIa samples following a
procedure that the data points will not be used again if they
have been matched to a QSO sample. The weighted average LD $\bar{D_{\rm L}}$ and its uncertainty $\sigma_{D_{\rm L}}$ can be obtained using conventional data processing techniques in Chapter 4 of Ref.~\cite{Bevington1993},
\begin{equation}
\label{avdi1}
\bar{D_{\rm L}}={\sum(D_{{\rm L}i}/\sigma_{D_{{\rm L}i}}^2)\over \sum1/\sigma_{D_{{\rm L}i}}^2},
\end{equation}
\begin{equation}
\label{erroravdi1}
\sigma^2_{\bar{D_{\rm L}}}={1\over \sum1/\sigma_{D_{{\rm L}i}}^2}\,.
\end{equation}
Here, $D_{{\rm L}i}$ denotes the $i$th appropriate luminosity distance data points, and $\sigma_{D_{{\rm L}i}}$ corresponds the observational uncertainty.

\begin{figure}[htbp]
\includegraphics[width=7.7cm]{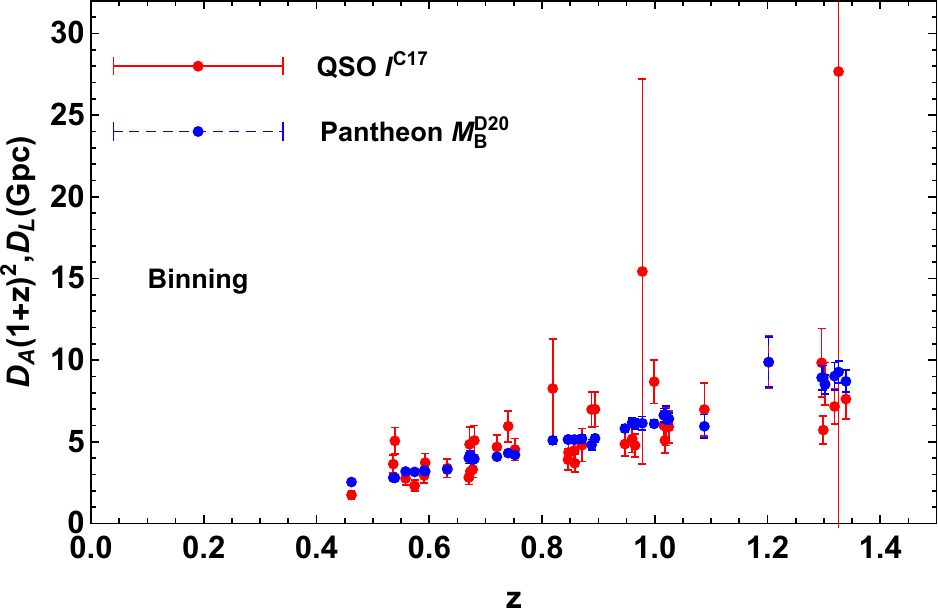}
\includegraphics[width=7.7cm]{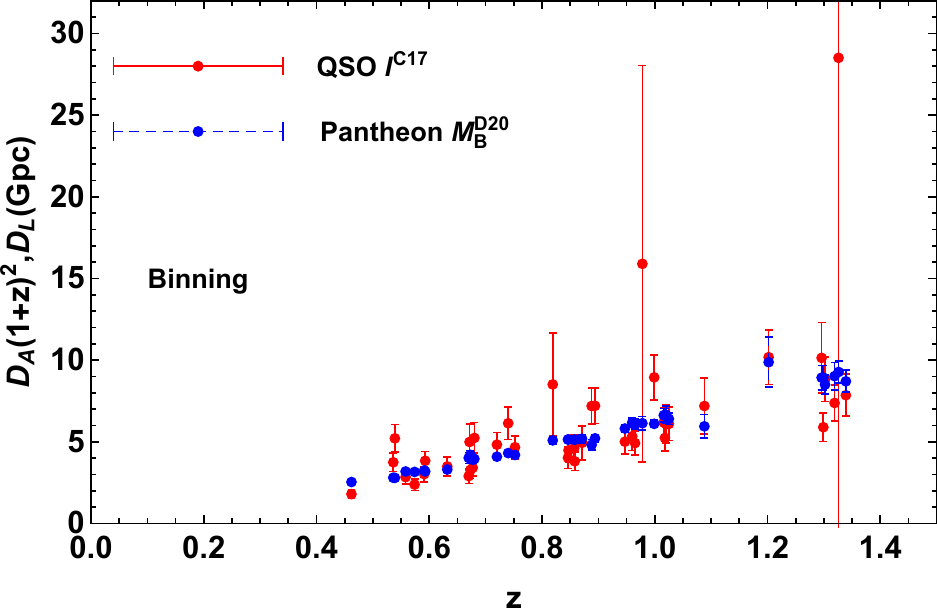}
\includegraphics[width=7.7cm]{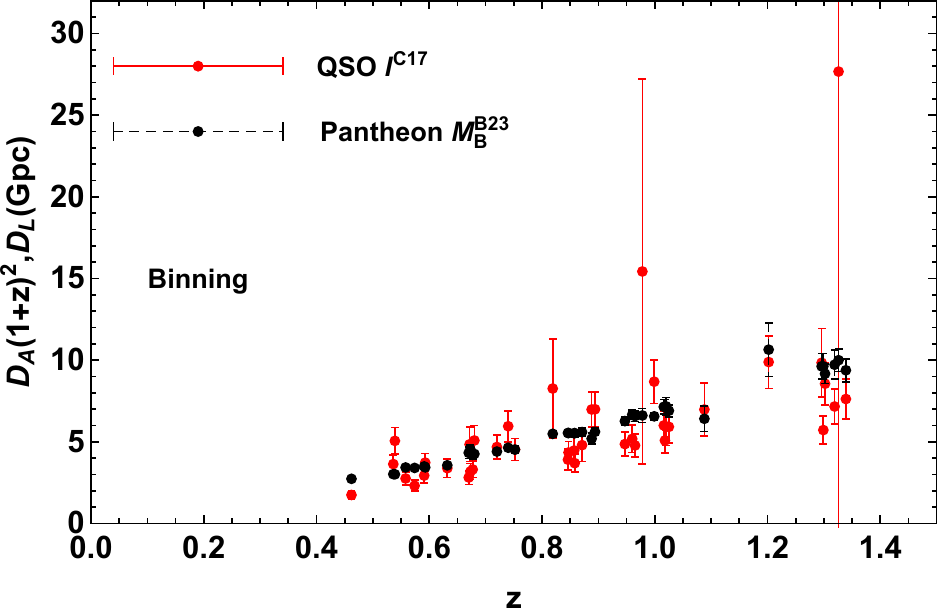}
\includegraphics[width=7.7cm]{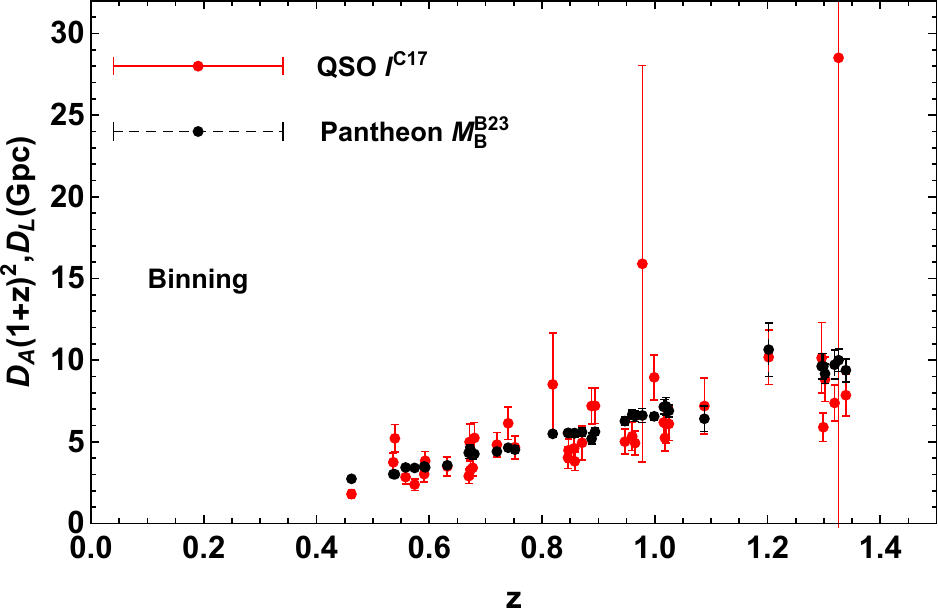}
\caption{\label{figdl1} In binning method, the sample catalogs of the observed  $D_{\rm A}(1+z)^2$ distribution from the QSO data points and the corresponding LD $D_{\rm L}$ from Pantheon data obtained with the priors of $M_{\rm B}^{\rm D20}$ (upper panel), $M_{\rm B}^{\rm B23}$ (bottom panel), $l^{\rm C17}$ (left panel) and $l^{\rm C19}$ (right panel), respectively.}
\end{figure}

Only 41 QSO data points satisfy the selection criterion. The distributions of the QSO data and SNIa data derived from different priors of $l$ and $M_{\rm B}$ are shown in Fig.~\ref{figdl1}.

\subsection{Artificial Neural Network}
It is important to note that when using selection criteria, one must be aware of the errors caused by the mismatch between SNIa and QSO data points. Additionally, most of the available QSO data points are excluded due to not meeting the selection criterion, as the density distribution of SNIa data differs from that of QSO data in certain redshift regions. To improve the robustness of QSO data when testing CDDR, we employ the ANN
to reconstruct the smoothing $m_{\rm B}(z)$ function from the Pantheon SNIa observations. Therefore, each ADD obtained from the QSO sample located within the redshift range of Pantheon SNIa has a corresponding LD of SNIa at the same redshift.

An ANN is usually a Deep Learning algorithm consisting of three layers: an input layer, a hidden layer, and an  output layer. The input layer comprises $n$ nodes, each of which corresponds to an independent variable, followed by $m$ interconnected hidden layers and the output layer with activation function nodes in the basic architecture~\cite{Schmidhuber2015}. ANN estimates the error gradient from observations in the training dataset, and
then updates the model weights and bias estimates during back propagation process to iterate toward an optimal
solution through the  Adam
optimization~\cite{Kingma2014}. The ANN process can be described  by vectorization representation, and more details can be found in Refs.~\cite{Wang2020,Clevert2015,LeCun2012}.

\begin{figure}[htbp]
\centering
\includegraphics[width=8cm]{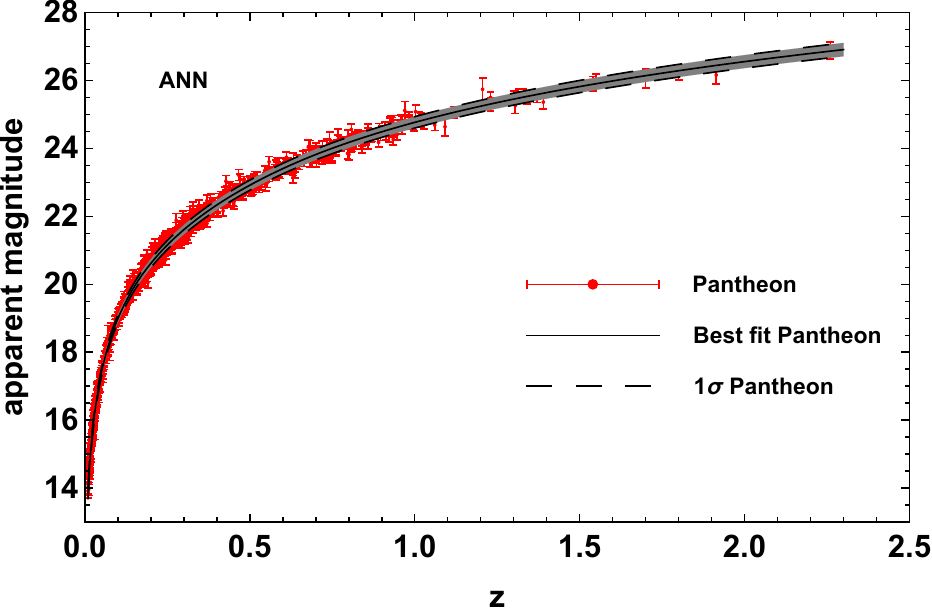}
\caption{\label{ANN} The distributions of the reconstructed $m_{\rm B}(z)$ function  with the corresponding $1\sigma$ errors with the ANN (black line), and the measurements of apparent magnitude from the Pantheon samples (red). }
\end{figure}

\begin{figure}[htbp]
\includegraphics[width=7.7cm]{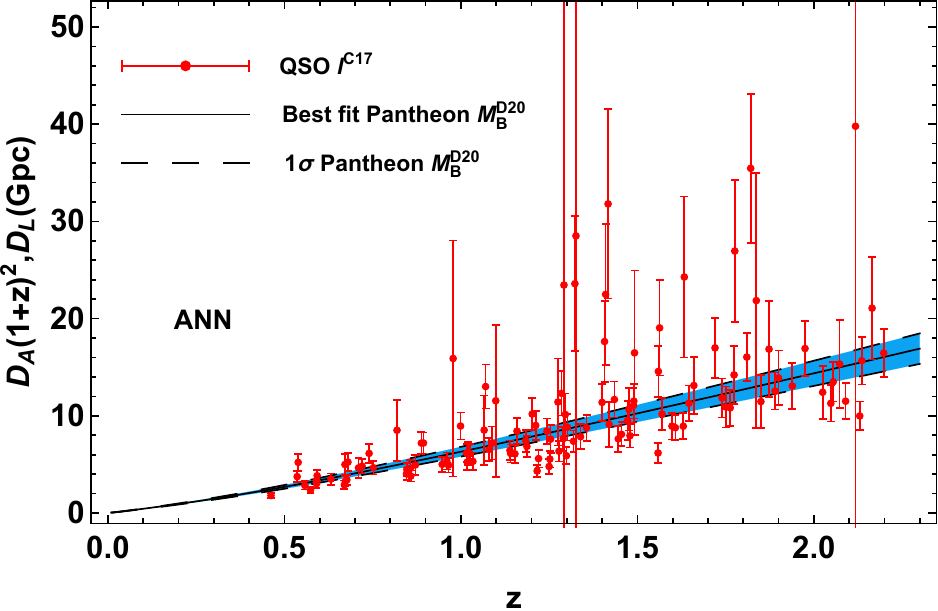}
\includegraphics[width=7.7cm]{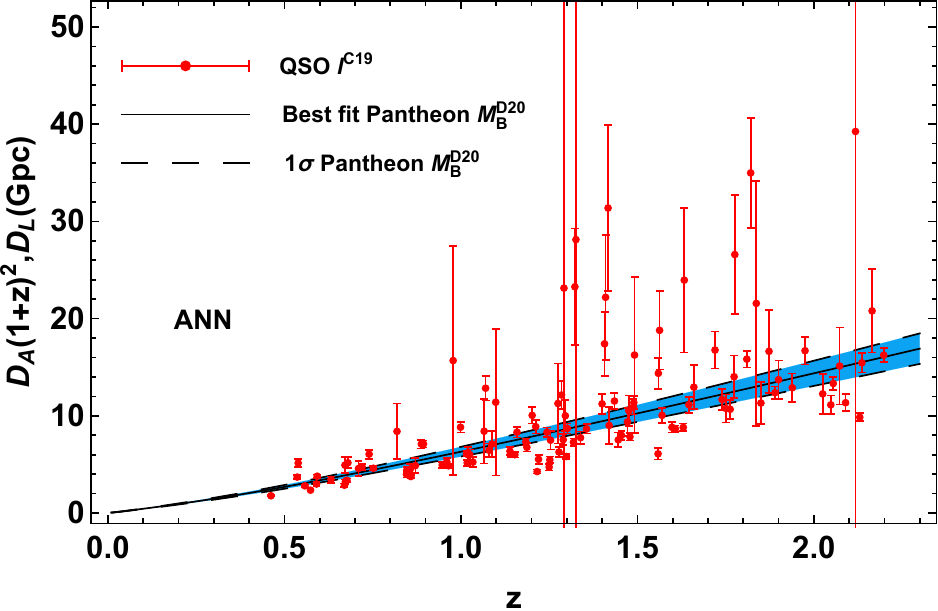}
\includegraphics[width=7.7cm]{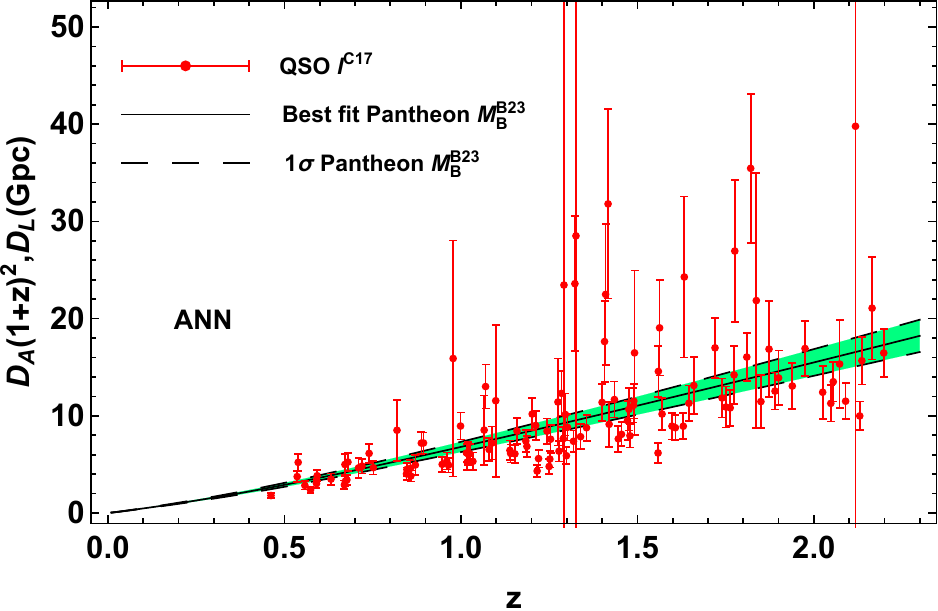}
\includegraphics[width=7.7cm]{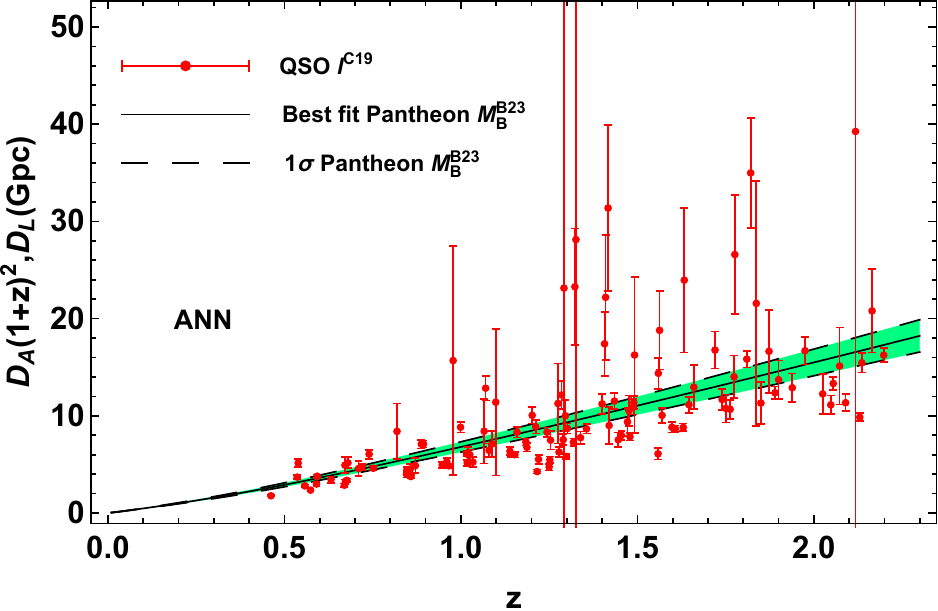}
\caption{\label{figdl2} The sample catalogs of the observed  $D_{\rm A}(1+z)^2$ distribution from the QSO data points derived with the priors of $l^{\rm C17}$ (left panel) and $l^{\rm C19}$ (right panel), and the LD $D_{\rm L}$ curves from Pantheon data derived with the priors of $M_{\rm B}^{\rm D20}$ (upper panel) and $M_{\rm B}^{\rm B23}$ (bottom panel).}
\end{figure}

We use the publicly available code, named Reconstructing Functions Using Artificial Neural Networks (ReFANN)\footnote{https://github.com/Guo-Jian-Wang/refann}~\cite{Wang2020}, to reconstruct the function of apparent magnitude $m_{\rm B}$ versus redshift $z$, as shown in Fig.~\ref{ANN}. It is easy to find that the uncertainty obtained from the ANN-reconstructed function are close to the observational uncertainty, and the reconstructed $1\sigma$ CL of the $m_{\rm B}$ can be considered as the average level of observational error.
The LD $D_{\rm L}$ corresponding to ADD $D_{\rm A}$ from QSO data points can be obtained through the smoothing function $m_{\rm B}(z)$ reconstructed by ANN. For the QSO samples, 116 QSO data points within the redshift range of SNIa observation $0 < z < 2.26$  can be matched with those from SNIa observation at the same redshift, and the remaining 4 QSO samples that are not within this redshift range are discarded. The distributions of the QSO data and reconstructed SNIa data derived from different priors of $M_{\rm B}$ and $l$ are shown in Fig.~\ref{figdl2}.

\subsection{Methodology}
We adopt the $\eta(z)$ function to verify any possible deviations from the CDDR at any redshift by comparing the $D_{\rm L}$ from SNIa and the $D_{\rm A}$ from QSO measurements. The $\eta(z)$ can be obtained through the following expression:
\begin{equation}\label{eta}
\eta(z)={D_{\rm L}\over D_{\rm A}}{(1+z)^{-2}}\,.
\end{equation}
At any redshift, $\eta(z)\neq 1$ indicates a deviation between the CDDR and astronomical observations.

\begin{figure}[htbp]
\includegraphics[width=7.7cm]{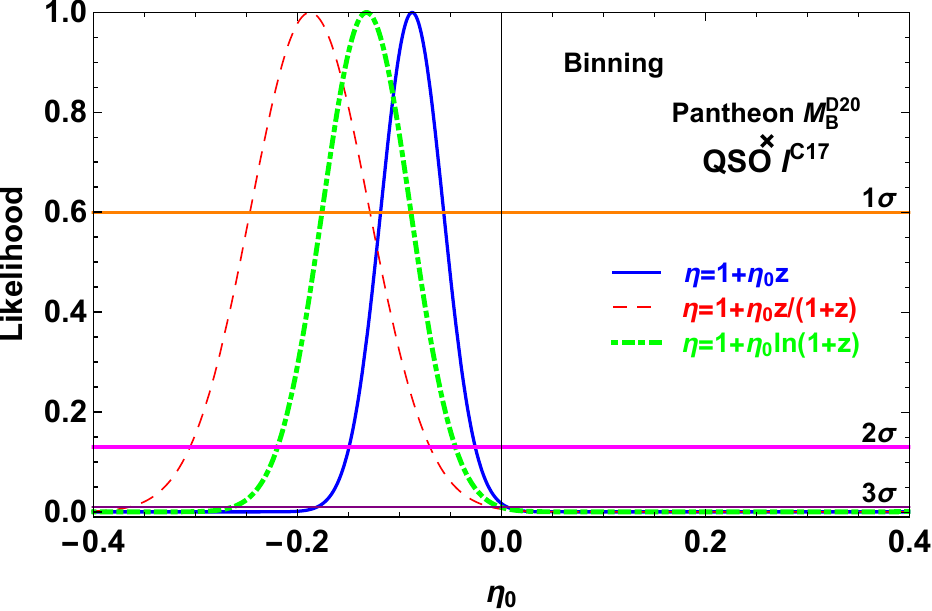}
\includegraphics[width=7.7cm]{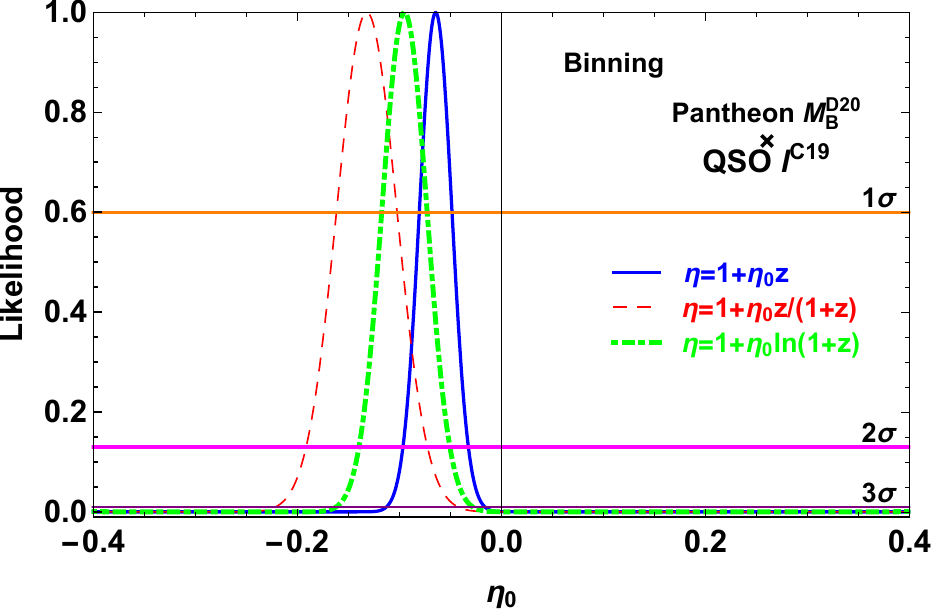}
\includegraphics[width=7.7cm]{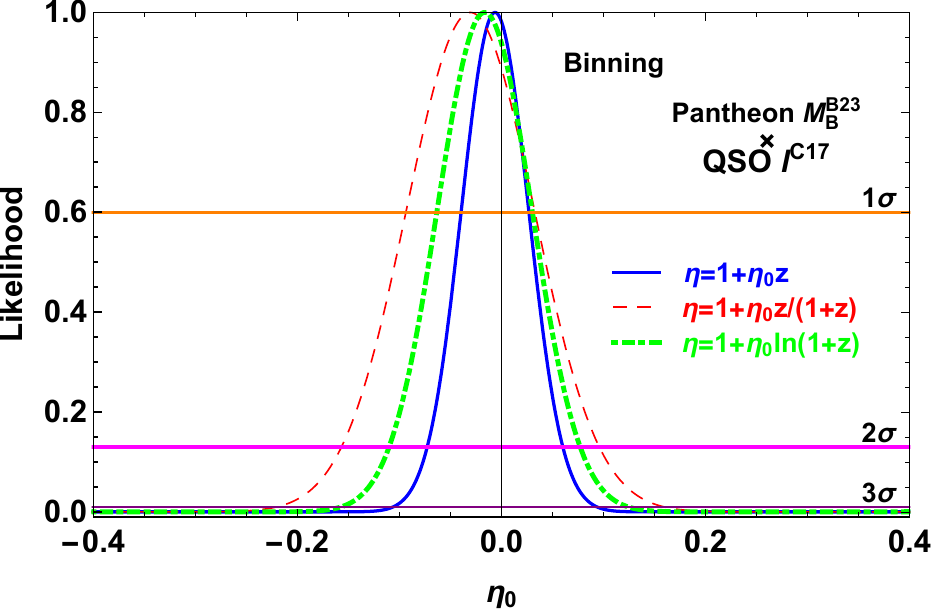}
\includegraphics[width=7.7cm]{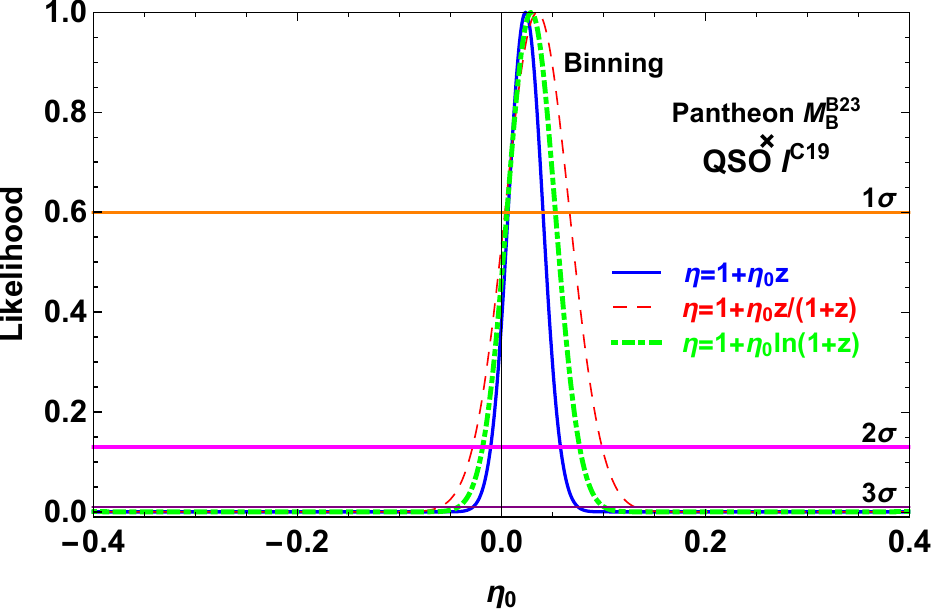}
\caption{\label{figetab1} In binning method, the likelihood distribution functions obtained with the priors of $M_{\rm B}^{\rm D20}$ (upper panel), $M_{\rm B}^{\rm B23}$  (bottom panel), $l^{\rm C17}$ (left panel) and $l^{\rm C19}$ (right panel). }
\end{figure}

\begin{figure}[htbp]
\includegraphics[width=7.7cm]{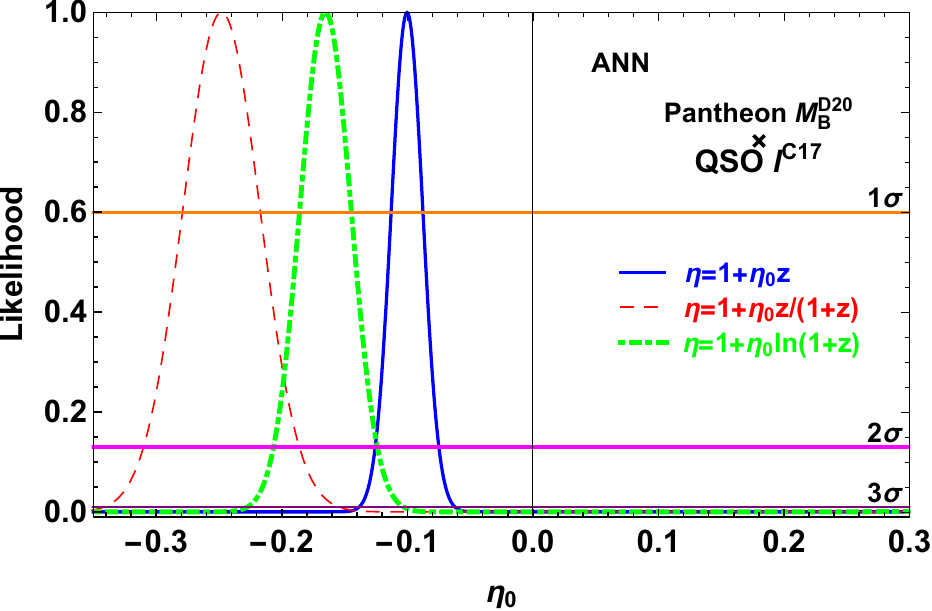}
\includegraphics[width=7.7cm]{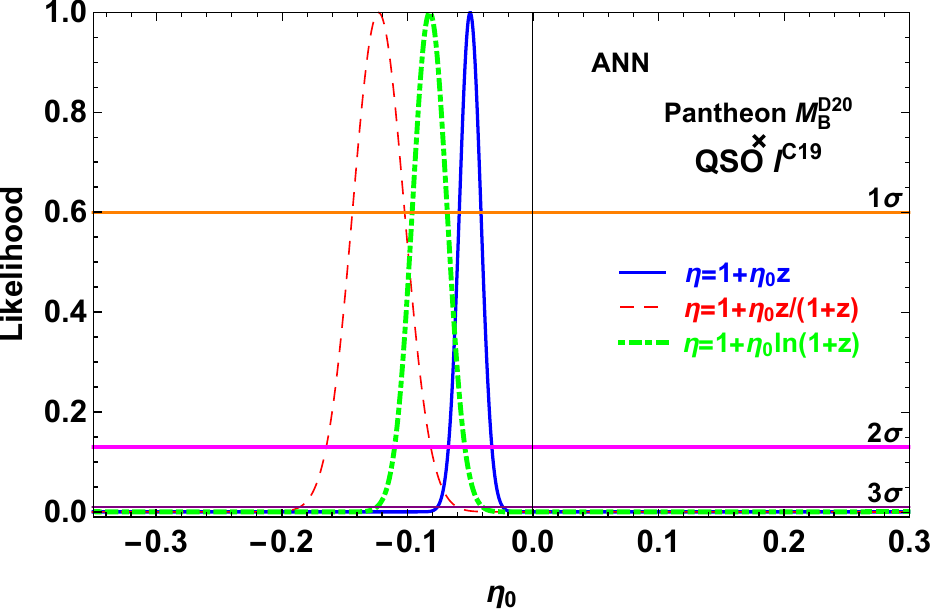}
\includegraphics[width=7.7cm]{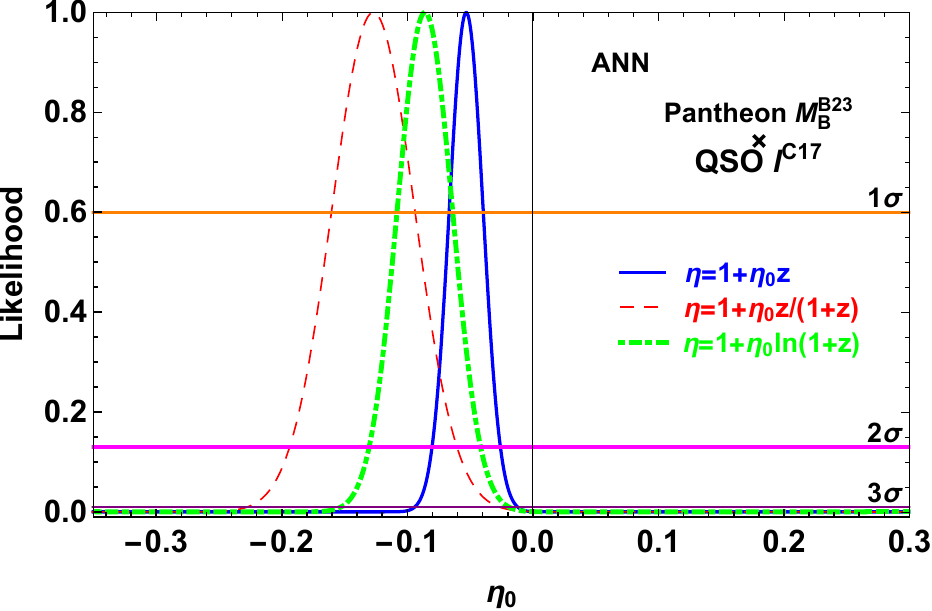}
\includegraphics[width=7.7cm]{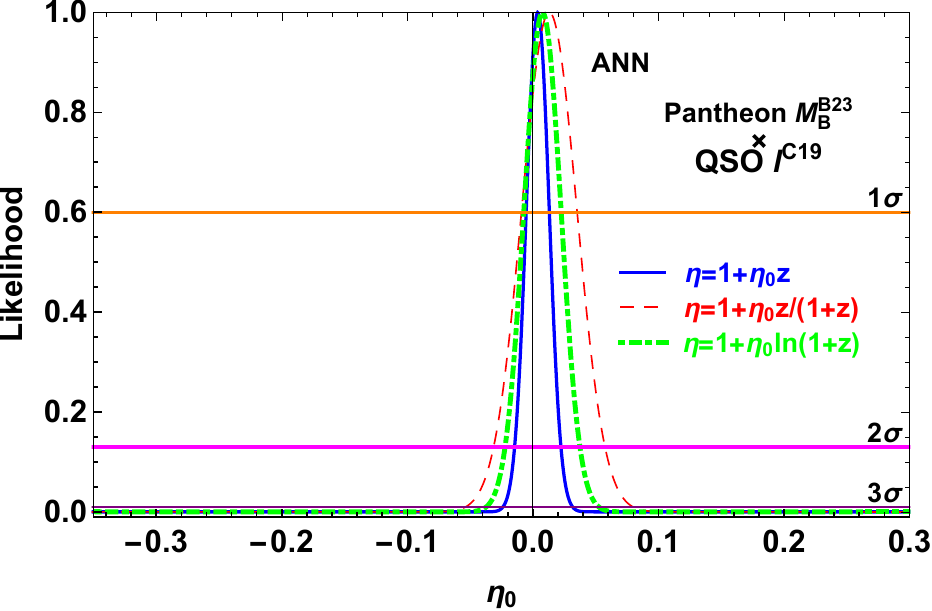}
\caption{\label{figetaA1} In ANN, the likelihood distribution functions obtained with the priors of $M_{\rm B}^{\rm D20}$ (upper panel), $M_{\rm B}^{\rm B23}$  (bottom panel), $l^{\rm C17}$ (left panel) and $l^{\rm C19}$ (right panel). }
\end{figure}

We adopt three types of parameterizations for $\eta(z)$: the linear form P1: $\eta(z)=1+\eta_0z$, and two non-linear forms P2: $\eta(z)=1+\eta_0z/(1+z)$, and P3: $\eta(z)=1+\eta_0\ln(1+z)$. The observed $\eta_{\rm obs} (z)$ is obtained from Eq.~\ref{eta}, and the corresponding error can be written as
\begin{equation}\label{SGL}
\sigma^2_{\eta_{\rm obs}}=\eta^2_{\rm obs}\left[\left({\sigma_{D_{\rm A}(z)}\over{D_{\rm A}(z)}}\right)^2+\left(\sigma_{D_{\rm L}(z)}
\over{D_{\rm L}(z)}\right)^2\right]\,.
\end{equation}
Thus,  we obtain
\begin{equation}
\label{chi}
\chi^{2}(\eta_0)=\sum_i^{N}\frac{{\left[\eta(z)-\eta_{{\rm obs},\,i}(z) \right] }^{2}}{\sigma^2_{\eta_{{\rm obs},i}}}\,.
\end{equation}
Here, $N$ represents the number of available QSO data points obtained the binning method or ANN. The constraint results on $\eta_0$ are shown in Fig.~\ref{figetab1}, Fig.~\ref{figetaA1}, and Tab.~\ref{likelihood1}. It is evident that the results obtained from the parametric method depend on the prior values of $M_{\rm B}$ or $l$. Thus, specific prior values of $M_{\rm B}$ or $l$ cause biases in the CDDR test, if their true values are not determined by astronomic observations.

Recently, Liu {\it et al.}~\cite{Liu2023a} used the fraction division $\eta(z_{i})/\eta(z_{j})$ to eliminate the impacts of $M_{\rm B}$ and $l$ on CDDR test, and the results indicated agreement between the CDDR and observations. More recently,  using the latest five BAO measurements and
the Pantheon SNIa sample, Xu {\it et al.}~\cite{Xu2022}  obtained CDDR test independently of the peak absolute magnitude $M_{\rm B}$ and  the sound horizon
scale $r_{\rm s}$ from transverse BAO measurements by marginalizing analytically the likelihood
function over the combination of $M_{\rm B}$ and  $r_{\rm s}$.
Since the uncertainty in an individual SNIa or QSO measurement is independent of $M_{\rm B}$ or $l$, these parameters can be removed from the fits by analytically marginalizing over them in the analysis. Following the process in Ref.~\cite{Xu2022}, we treat the fiducial values of $M_{\rm B}$ and $l$ as nuisance parameters to determine the LD $D_{\rm L}$ and ADD $D_{\rm A}$, and then marginalize their effect by using a flat prior in the statistic analysis. The likelihood distribution $\chi^{\prime\,2}$ can be rewritten as
\begin{equation}
\label{chi2}
\chi^{\prime\,2}(\eta_0, \kappa)= \sum_i^{N}\frac{{{\alpha_i^2 \over \beta_i^2}{\kappa}^2- 2 {\alpha_i \over \beta_i}{\kappa}+1  }}{\sigma^{\prime\,2}_{{\eta_{{\rm obs},i}}}}\,,
\end{equation}

Here, $\alpha_i=\eta(z_i)$, $\beta_i=10^{({m_{{\rm B},i}\over 5}-5)}\theta_{{\rm QSO},i}(1+z_i)^{-2}$, $\kappa=(10^{M_{\rm B} \over 5}l)$, and
\begin{equation}
\label{sigma01}
\sigma_{\eta_{\rm obs},i}^{\prime\,2}=\left({\ln{10}\over {5}}{\sigma_{m_{{\rm B},i}}}\right)^2+\left({\sigma_{\theta_{{\rm QSO},i}}\over{\theta_{{\rm QSO},i}}}\right)^2\,.
\end{equation}

Thus, following the approach described in Refs.~\cite{Xu2022,Wang2023,Conley2011}, the marginalized $\chi^{\prime\,2}$ in Eq.~\ref{chi2} can be rewritten as:

\begin{equation}
\label{chi3}
\chi_{\rm M}^{\prime\,2}(\eta_0)= C-{B^2\over {A}}+\ln{A\over 2\pi}\,,
\end{equation}
where $A=\sum \alpha_i^2/(\beta_i^2{\sigma^{\prime\,2}_{{\eta_{{\rm obs},i}}}})$, $B=\sum \alpha_i/(\beta_i{\sigma^{\prime\,2}_{{\eta_{{\rm obs},i}}}})$, and $C=\sum 1/{\sigma^{\prime\,2}_{{\eta_{{\rm obs},i}}}}$.

\begin{figure}[htbp]
\includegraphics[width=7.7cm]{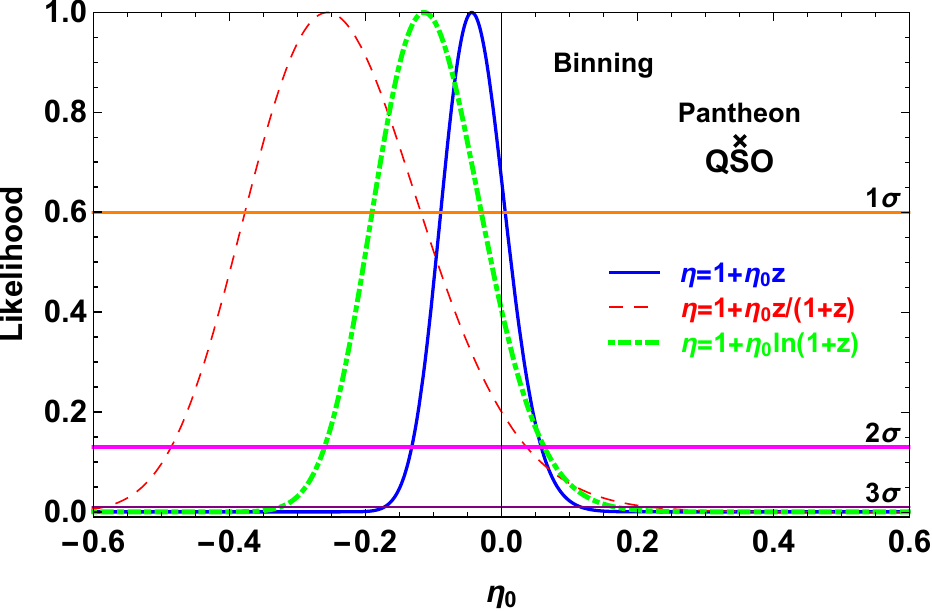}
\includegraphics[width=7.7cm]{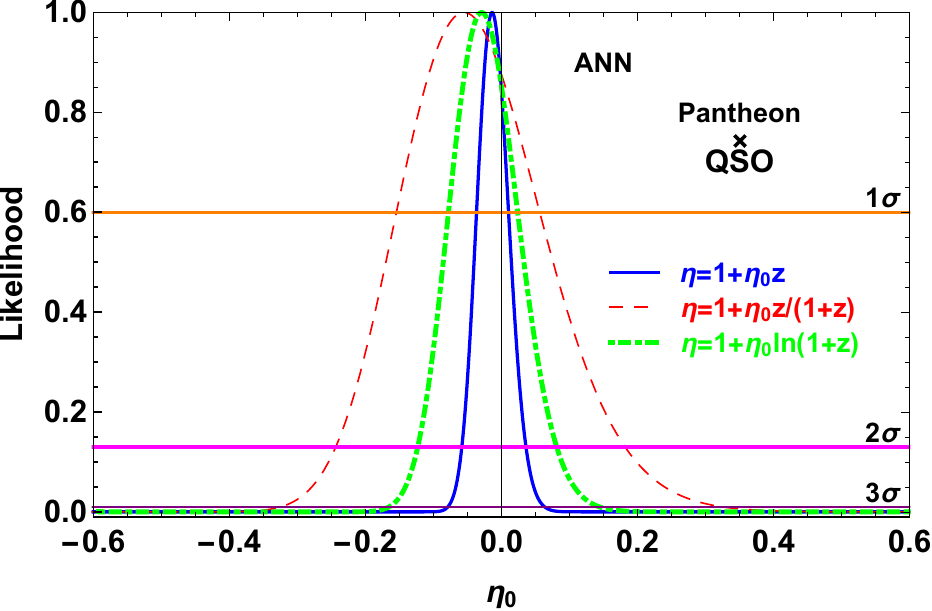}
\caption{\label{figetabA} The likelihood distribution obtained with flat priors on $\kappa$ using the binning method (left) and ANN (right). }
\end{figure}

It is evident that all of the quantities used in the CDDR test come directly from observations, and $\chi_{\rm M}^{\prime\,2}$ in Eq.~\ref{chi3} is independent of parameters such as $M_{\rm B}$ and $l$. In this way, we can remove $M_{\rm B}$ and $l$ from the fit by analytically marginalizing them in Eq.~\ref{chi2}. Since this test is based on the observed data and does not require any assumptions about cosmological models, the parametric method used to test CDDR is independent of the cosmological model. The results are shown in Fig.~\ref{figetabA} and Tab.~\ref{likelihood1}. To compare the capability of QSO data with that of other astronomic observational data in testing CDDR, we list the results of the constraints on $\eta_0$ obtained from different observational data sets in Tab.~\ref{otherobser}.

\begin{table}[htp]
\begin{tabular}{|c|c|c|c|c|}
\hline
\scriptsize{parmetrization }  & \   P1: $1+\eta_0 {z}$\ \ &P2: $1+\eta_0 {z\over(1+z)}$  & \   P3: $1+\eta_0 {\ln(1+z)}$ \\
\hline
\scriptsize{$\eta_0^{\rm {A\,C}\dag}$} & \scriptsize{${-0.088\pm0.031\pm0.062\pm0.093}$}& \scriptsize{${-0.188\pm0.059\pm0.118\pm0.177}$} & \scriptsize{${-0.132\pm0.044\pm0.087\pm0.131}$} \\
\hline
\scriptsize{$\eta_0^{\rm {A\,C}\ddag}$} & \scriptsize{${-0.100\pm0.013\pm0.025\pm0.038}$}& \scriptsize{${-0.248\pm0.031\pm0.062\pm0.093}$} & \scriptsize{${-0.165\pm0.021\pm0.041\pm0.062}$} \\
\hline
\scriptsize{$\eta_0^{\rm {B\,C}\dag}$}  & \scriptsize{${-0.006\pm0.033\pm0.066\pm0.100}$}& \scriptsize{${-0.030\pm0.063\pm0.126\pm0.190}$} & \scriptsize{${-0.017\pm0.047\pm0.094\pm0.141}$} \\
\hline
\scriptsize{$\eta_0^{\rm {B\,C}\ddag}$}  & \scriptsize{${-0.053\pm0.014\pm0.027\pm0.041}$}& \scriptsize{${-0.127\pm0.033\pm0.067\pm0.100}$} & \scriptsize{${-0.086\pm0.022\pm0.044\pm0.067}$} \\
\hline
\scriptsize{$\eta_0^{\rm {A\,D}\dag}$} & \scriptsize{${-0.065\pm0.016\pm0.032\pm0.048}$}& \scriptsize{${-0.132\pm0.030\pm0.059\pm0.089}$} & \scriptsize{${-0.095\pm0.022\pm0.045\pm0.067}$} \\
\hline
\scriptsize{$\eta_0^{\rm {A\,D}\ddag}$} & \scriptsize{${-0.050\pm0.009\pm0.017\pm0.026}$}& \scriptsize{${-0.123\pm0.021\pm0.042\pm0.063}$} & \scriptsize{${-0.082\pm0.014\pm0.028\pm0.042}$} \\
\hline
\scriptsize{$\eta_0^{\rm {B\,D}\dag}$} & \scriptsize{${0.024\pm0.017\pm0.034\pm0.051}$}& \scriptsize{${0.035\pm0.032\pm0.063\pm0.095}$} & \scriptsize{${0.029\pm0.024\pm0.048\pm0.071}$} \\
\hline
\scriptsize{$\eta_0^{\rm {B\,D}\ddag}$} & \scriptsize{${0.004\pm0.009\pm0.018\pm0.028}$}& \scriptsize{${0.013\pm0.022\pm0.044\pm0.067}$} & \scriptsize{${0.008\pm0.015\pm0.030\pm0.045}$} \\
\hline
\scriptsize{${\eta_0}^{\star\dag}$} & \scriptsize{${-0.044\pm^{0.049}_{0.046}\pm^{0.102}_{0.088}\pm^{0.160}_{0.128}}$}& \scriptsize{${-0.256\pm^{0.137}_{0.121}\pm^{0.292}_{0.230}\pm^{0.472}_{0.328}}$} & \scriptsize{${-0.114\pm^{0.084}_{0.076}\pm^{0.176}_{0.146}\pm^{0.280}_{0.211}}$} \\
\hline
\scriptsize{${\eta_0}^{\star\ddag}$} & \scriptsize{${-0.014\pm^{0.024}_{0.023}\pm^{0.050}_{0.044}\pm^{0.077}_{0.065}}$}& \scriptsize{${-0.055\pm^{0.111}_{0.099}\pm^{0.237}_{0.189}\pm^{0.380}_{0.270}}$} & \scriptsize{${-0.029\pm^{0.053}_{0.049}\pm^{0.110}_{0.094}\pm^{0.173}_{0.136}}$} \\
\hline
\end{tabular}
\caption{The maximum likelihood estimation results for the parameterizations with the binning method and ANN. The $\eta_0$ is represented by the best fit value $\zeta_{0,{\rm best}} \pm 1\sigma \pm 2\sigma \pm 3\sigma$ for each dataset. The superscripts A, B, C, and D represent the cases obtained from $M_{\rm B}^{\rm D20}$, $M_{\rm B}^{\rm B23}$, $l^{\rm C17}$, and $l^{\rm C19}$, respectively. The superscript $\star$ denotes the results obtained from the flat marginalization for $M_{\rm B}$ and $l$, and $\dag$ and $\ddag$ denote the results obtained from the binning method and ANN, respectively. }
\label{likelihood1}
\end{table}

\begin{table*}[htb]
\label{tab:results}
\begin{center}
\begin{tabular}{ccccc} \hline \hline
Dataset used & P1: $1+\eta_0 {z}$\ \ &P2: $1+\eta_0 {z\over(1+z)}$  & \   P3: $1+\eta_0 {\ln(1+z)}$ \\ \hline
${\rm {Union2 }}+{\rm {Galaxy(Prior)}}$~\cite{Li2011} & ${-0.22{\pm{0.11}}}$ & ${-0.33{\pm{0.16}}}$    \\ [1ex]
${\rm {Union2.1}}+{\rm {91GMF(Prior)}}$~\cite{Goncalves2015} & ${-0.08{\pm^{0.11}_{0.10}}}$    \\ [1ex]
${\rm {Union2.1}}+{\rm {BAO(Prior)}}$~\cite{Wu2015} & ${-0.086{\pm{0.064}}}$ &  ${-0.131{\pm{0.098}}}$    \\ [1ex]
${\rm {Union2.1}}+{\rm {BAO(Marg)}}$~\cite{Wu2015} & ${-0.174{\pm^{0.253}_{0.199}}}$ & ${-0.409{\pm^{0.529}_{0.381}}}$    \\ [1ex]
${\rm {Y_{SZ}}}-{\rm {Y_{X}}}\, {\rm {ratio}}+H(z)(\rm Prior)$~\cite{Bora2021} & ${0.008{\pm{0.05}}}$ & ${0.019{\pm{0.11}}}$ & ${0.013{\pm{0.07}}}$   \\ [1ex]
${\rm {GMF}}+{\rm {SNIa}}+{T_{\rm CMB}}(\rm Prior)$~\cite{Holanda2017} & ${-0.020{\pm{0.027}}}$ &  ${-0.041{\pm{0.042}}}$    \\ [1ex]
${\rm {SNIa }}+{\rm {BAO}(Marg)}$~\cite{Xu2020} & ${-0.07{\pm{0.12}}}$ & ${-0.20{\pm{0.27}}}$  & ${-0.12{\pm{0.18}}}$  \\ [1ex]
${\rm {SNIa }}+{\rm {BAO}(Marg)}$~\cite{Xu2022} & ${-0.037{\pm^{0.110}_{0.097}}}$ & ${-0.101{\pm^{0.269}_{0.225}}}$  & ${-0.061{\pm^{0.173}_{0.149}}}$  \\ [1ex]
\hline \hline
\end{tabular}
\caption[]{Summary of the constraints on parameter $\eta_0$ with different data sets. ``Prior'' represents the results obtained using certain parameters with specific priors, and ``Marg'' represents the results obtained by marginalizing certain parameters with a flat prior.}
\label{otherobser}
\end{center}
\end{table*}

\section{ Results and Analysis}
From Fig.~\ref{figetab1} and Tab.~\ref{likelihood1}, it can be seen that through the binning method,  CDDR is consistent with the observed data at various confidence levels (CLs) depending on the combination of $M_{\rm B}$ and $l$. Specifically, it is consistent at $3\sigma$ CL  with the combination $M_{\rm B}^{\rm D20}$ and $l^{\rm C17}$; at $1\sigma$ CL with the combination of $M_{\rm B}^{\rm B23}$ and $l^{\rm C17}$; and at $2\sigma$ CL with the combination of $M_{\rm B}^{\rm B23}$ and $l^{\rm C19}$. However, CDDR is not consistent with the observed data at $3\sigma$ CL with the combination of $M_{\rm B}^{\rm D20}$ and $l^{\rm C19}$. Similarly, from Fig.~\ref{figetaA1} and Tab.~\ref{likelihood1}, through the ANN method, CDDR is only consistent with the observed data at $1\sigma$ CL with the combination of $M_{\rm B}^{\rm B23}$ and $l^{\rm C19}$. Nevertheless, when considering the other combinations of $M_{\rm B}$ and $l$ values, specifically $M_{\rm B}^{\rm B23}$ and $l^{\rm C17}$, $M_{\rm B}^{\rm D20}$ and $l^{\rm C17}$, as well as $M_{\rm B}^{\rm D20}$ and $l^{\rm C19}$, the CDDR is not consistent with the observed data. Therefore, the prior values of $M_{\rm B}$ and $l$ may induce significant bias in the CDDR test.

From Fig.~\ref{figetab1}, Fig.~\ref{figetaA1}, and Tab.~\ref{likelihood1}, the parametrization P1 imposes the most rigorous constraints on testing CDDR, although the result of the CDDR test is nearly independent of the parametrization of $\eta(z)$. Now, we compare the capability of QSO measurements to constrain parameter $\eta_0$ with that of other astronomical observations obtained under specific prior conditions of cosmological variables. With the binning method, the QSO measurements improve the accuracy of $\eta_0$ by approximately 85\% at $1\sigma$ CL when compared to results obtained from the Union2+galaxy cluster observation (the elliptical $\beta$ model)~\cite{Li2011} or Union2.1+91gas mass fraction (GMF) observation~\cite{Goncalves2015}; and about 75\% at $1\sigma$ CL when compared to results obtained from Union2.1+BAO measurements~\cite{Wu2015}, where the CDDR tests were conducted with specific priors of $M_{\rm B}$ or $r_{\rm s}$ from the CMB observations. Our results are also roughly 60\% more stringent than those from the South Pole Telescope-SZ clusters and X-ray measurements from Multi-mirror Mission-Newton~\cite{Bora2021}, where the prior of $M_{\rm B}$ and $H_0$ are utilized; and are 40\% more stringent than the result from the x-ray GMF of galaxy clusters jointly with SNIa and CMB temperature~\cite{Holanda2017}, where $M_{\rm B}$ is fixed to derive the LD. Using the ANN, while  testing the CDDR with more available QSO measurements, the uncertainties of $\eta_0$ at the $1\sigma$ CL are improved by approximately 50\%  when compared to results obtained from the binning method.

When testing the CDDR using a flat prior of $\kappa\equiv 10^{M_{\rm B}\over 5}l$, CDDR is consistent with the observed data at $2\sigma$ CL with the binning method; and at $1\sigma$ CL with the ANN method. The constraints on $\eta_0$ obtained from the flat prior of $\kappa$  are much weaker than those obtained from the specific priors of $M_{\rm B}$ and $l$, due to marginalizing $\kappa$ with a flat prior in our analysis. To assess the ability of testing CDDR from QSO measurements, it is valuable to compare our results with previous constraints on $\eta_0$ from other observational data by marginalizing certain parameters with a flat prior. With the binning method, the QSO measurements improve the accuracy of $\eta_0$ by approximately 75\% at $1\sigma$ CL when compared to the results obtained from the Union2.1+BAO observations, where the dimensionless Hubble constant $h$ was marginalized with a flat prior~\cite{Wu2015}. The constraints on $\eta_0$ are roughly 60\% more stringent than those from the most recent Pantheon samples and BOSS DR12 BAO measurements within the redshift range $0.31\leq{z}\leq0.72$~\cite{Xu2020}; and they are 50\% more stringent than the result from 5 BAO measurements utilizing the extended Baryon Oscillation Spectroscopic Survey data release 16 quasar samples in conjunction with the Pantheon SNIa samples~\cite{Xu2022}, where $M_{\rm B}$ and $r_{\rm s}$ were marginalized. As for the results obtained from the ANN method, the uncertainties of $\eta_0$ at $1\sigma$ CL are reduced by approximately 40\%  when compared to the results obtained from the binning method. Therefore, the QSO measurements demonstrate a superior capability in testing CDDR compared to BAO observations, which have been recognized as powerful tools for testing CDDR~\cite{Wu2015,Xu2020}.

\section{conclusion}
The CDDR plays an important role in astronomical observations and modern cosmology, and any deviation from the CDDR may indicate new physical signals. SNIa and QSO measurements can be regarded as effective observational data for testing the CDDR. However, due to the uncertainty in the absolute magnitude $M_{\rm B}$ and the linear size scaling factor $l$ , which are constrained by different astronomical observations and cosmological models, it is necessary to investigate the impact of the prior values of $M_{\rm B}$ and $l$ on the CDDR test, and to verify the validity of the CDDR using new methods.

In this work, we test the CDDR by comparing the LD derived from the Pantheon SNIa compilation with ADD from QSO measurements, using parametric methods. We employ the binning method and ANN to match the SNIa data with the QSO measurements at the same redshift, and adopt the function $\eta(z)=D_{\rm L}(z)/D_{\rm A}(z)(1 + z)^{-2}$ to probe the possible deviations from the CDDR at any redshift. Two specific prior values of  $M_{\rm B}$ and  $l$ are used to obtain the LDs from the SNIa observations and the ADDs from QSO measurements, respectively.  The results show that the specific prior values of $M_{\rm B}$ and $l$ cause significant biases in the CDDR test, if the astronomical observations do not provide accurate values for $M_{\rm B}$ and $l$.

To avoid the bias in the CDDR test caused by the prior values of $M_{\rm B}$ and $l$, we treat the fiducial values of $M_{\rm B}$ and $l$ as nuisance parameters to determine the LD ${D_{\rm L}}$ and ADD ${D_{\rm A}}$. We then marginalize their impacts on the CDDR test by applying a flat prior on the new variable $\kappa$ $\equiv$ $10^{M_{\rm B}\over 5}l$ in the statistical analysis. Our results indicate no violation of the CDDR. However, the capability of the QSO measurements to test CDDR is reduced compared to the results obtained from specific values of $M_{\rm B}$ and $l$, due to marginalizing $\kappa$ with a flat prior in our analysis. In comparison to the previous results, the capability of QSO measurements to test the CDDR is much stronger than that of other previous astronomic observations, regardless of whether the method used is dependent on $M_{\rm B}$ and $l$ or not. It is noteworthy that the method for testing the CDDR in this work is not only independent of the cosmological model but also independent of the prior values of the absolute magnitude $M_{\rm B}$ and the linear size scaling factor $l$. Therefore, QSO measurement can serve as a powerful tool for testing CDDR independently of cosmological model.

\begin{acknowledgments}
This work was supported by the National Natural Science Foundation of China under Grants No. 12375045,
No. 12305056, No. 12105097 and No. 12205093, the
Hunan Provincial Natural Science Foundation of China
under Grants No. 12JJA001 and No. 2020JJ4284, the Natural Science Research Project of Education
Department of Anhui Province No. 2022AH051634, and the
Science Research Fund of Hunan Provincial Education
Department No. 21A0297.

\end{acknowledgments}
\bibliography{refs}

\end{document}